\documentclass[prb,aps,twocolumn,showpacs,amsmath,amssymb,superscriptaddress]{revtex4}
\usepackage{epsfig}
\usepackage{graphicx}
\usepackage{dcolumn}
\usepackage{bm}
\begin{document}
\title{Evolution of electrical resistivity and electronic structure 
of transuranium metals under pressure: Numerical investigations}
\author{Yu.~Yu.~Tsiovkin}
\email{y.tsiovkin@mail.ustu.ru}
\affiliation{Ural Federal University,
620002 Yekaterinburg, Russia}
\author{A.~V.~Lukoyanov}
\affiliation{Ural Federal University,
620002 Yekaterinburg, Russia}
\affiliation{Institute of Metal Physics, Russian Academy of Sciences--Ural Division,
620990 Yekaterinburg, Russia}
\author{A.~O.~Shorikov}
\affiliation{Ural Federal University,
620002 Yekaterinburg, Russia}
\affiliation{Institute of Metal Physics, Russian Academy of Sciences--Ural Division,
620990 Yekaterinburg, Russia}
\author{A.~A.~Povzner}
\affiliation{Ural Federal University,
620002 Yekaterinburg, Russia}
\author{L.~Yu.~Tsiovkina}
\affiliation{Ural Federal University,
620002 Yekaterinburg, Russia}
\author{A.~A.~Dyachenko}
\affiliation{Ural Federal University,
620002 Yekaterinburg, Russia}
\author{V.~B.~Bystrushkin}
\affiliation{Ural Federal University,
620002 Yekaterinburg, Russia}
\author{M.~V.~Raybukhina}
\affiliation{Ural Federal University,
620002 Yekaterinburg, Russia}
\author{M.~A.~Korotin}
\affiliation{Institute of Metal Physics, Russian Academy of Sciences--Ural Division,
620990 Yekaterinburg, Russia}
\author{V.~V.~Dremov}
\affiliation{Russian Federal Nuclear Center, Institute of Technical Physics, Snezhinsk, 
456770 Chelyabinsk Region, Russia}
\author{V.~I.~Anisimov}
\affiliation{Institute of Metal Physics, Russian Academy of Sciences--Ural Division,
620990 Yekaterinburg, Russia}

\date{\today}
\begin{abstract}

The influence of electronic structure evolution upon pressure on the temperature dependencies 
of electrical resistivity of pure Np, Pu, Am, and Cm metals have been investigated 
within coherent potential approximation (CPA) for many-bands conductivity model. 
Electronic structure of pure actinide metals was calculated within the local density approximation 
with the Hubbard $U$ and spin-orbit coupling corrections (LDA+$U$+SO) method in various phases 
at normal conditions and under pressure. They were compared with the corresponding 
cubic (bcc or fcc) phases. The densities of states of the latter were used as a starting point 
of model investigations of electrical resistivity. The obtained results were found 
in good agreement with available experimental data. The nature of large magnitude of resistivity 
of actinides was discussed in terms of the proposed conductivity model using 
the $ab~initio$ calculated parameters.

\end{abstract}


\maketitle
\section{Introduction}

For many decades transuranium metals have been of interest to scientific community 
due to the unique combination of structural, electronic, and kinetic properties.~\cite{LosAlamos} 
Their rich phase diagrams\cite{Young01} include a number of structural 
transitions upon pressure or increasing the temperature.~\cite{HeckerPMS,Lander03,Heathman01,Lindbaum01,Heathman05} 
Intermediate or $jj$ coupling schemes, in contrast to usual Russel-Saunders ($LS$) 
one, were found to be more appropriate for 5$f$ electrons in transuranium metals 
due to the presence of large spin-orbit coupling in actinides.~\cite{Moore09,Moore03,Laan04} 
Additionally, magnetic stabilization of phases was found 
in curium.~\cite{Heathman05,Moore07} Drastic changes of crystal and electronic structures 
and their interplay\cite{Migl10} do not deplete this list of unusual properties of the metals. 

Anomalous resistance properties in transuranium metals are also well known.~\cite{LA} 
The negative temperature coefficient of resistivity (TCR) was measured in $\alpha$- 
and $\delta$-Pu  and some Pu-based alloys at high temperatures ($T>\theta_{D}$).~\cite{Sm,Oll} 
Moreover, $\rho \sim T^2$ dependence was found in these compounds at low temperature.~\cite{Brod,Brodsky,LA} 
Electrical resistivity (ER) behavior in pure Np and Am metals is ordinary,~\cite{ Np_res1,Np_res2,AmR} 
namely, $\rho \sim T $ in high and $\rho \sim T^3$ and $\rho \sim T^{4\div5}$ 
in low temperature ranges, respectively.~\cite{remark} The superconducting state 
was found in Am at ambient pressure with the characteristic temperature 
T$_c$~$\sim$~0.5 -- 2.0~K.~\cite{B} Curium metal also demonstrates 
ordinary behavior of ER typical for antiferromagnetic metal with $\rho \sim T^2 $ 
at low temperature and $\rho \sim T $ at high temperature, respectively. 
However, temperature dependencies of ER in transuranium metals show 
anomalously high values of ER (typically, $80 \div 140~ \mu k\Omega \cdot cm $) 
at room temperature. For comparison, ER of transition metals 
at the same conditions are about $1 \div 10~ \mu k \Omega \cdot cm$.

Such high resistivity values of trasuranium metals have no reasonable explanation yet 
and all previous model investigations substantially underestimate the experimental 
values of ER. For example, Mott two-band conductivity model combined with 
coherent potential approximation (CPA) shows, that Np, Pu, Am, and Cm are typical 
metals at high temperatures,~\cite{Tsiovkin07} 
and observed ER behavior is described well in relative units. However, absolute values 
of ER in transuranium metals cannot be estimated from these calculations.
Indeed, the estimation of maximum value of ER -- taking into account only interband transitions 
of scattering conductivity electrons with additional accounting that probabilities are 
proportional to the DOSes values at the Fermi level -- gives the ER values at most 
$20 \div 40~ \mu k\Omega \cdot cm$ that underestimates
the experimental data by a factor~of~3.

It is well known, that the high-resistivity values can be provided by specific Kondo-like resonance. 
Actinides, usually considered as heavy-fermion systems,~\cite{Moore09} should 
demonstrate appropriate kinetic and magnetic properties in low temperature range. However, 
we have no experimental data for the detailed discussion in terms of such a model. 
Previously reported investigations of ER within common Kondo model\cite{Me, Dall} have a number 
of problems and ambiguities\cite{Fluss} due to selecting of the ``magnetic'' part of ER.  
A principal problem arose if one uses spin-fluctuation model\cite{Ju} for describing 
ER of $\delta$-Pu and other actinides. Indeed, one can find good agreement of calculated 
and experimental data only adjecting the Stoner factor.~\cite{Brod} However, such a high magnitude 
of the Stoner factor should result in specific temperature behavior of magnetic 
susceptibility, typical for amplified paramagnets. In contrast to these expectations, 
ordinary for paramagnets temperature dependence of magnetic susceptibility was measured 
in the systems under consideration.~\cite{Me}

On the other hand, recent NMR experiments for $\delta$-Pu revealed a combination 
of Curie-Weiss and van-Fleck temperature behavior of spin susceptibility.~\cite{Stas1,Stas2} 
That demonstrates strong influence of spin-density fluctuations on electronic structure 
and can provide anomalies of electron heat capacity in the same temperature range, 
as reported in Ref.~\onlinecite{LashlyPRL}. Actually, strong  spin-density fluctuations 
effects can play a noticeable role in actinides, providing $\rho \sim T^2$ in very low 
temperature region, and suppressing superconductivity. Note also, that the $\rho \sim T^2$ 
dependence and negative TCR (observed in $\delta$-Pu) can be described in terms 
of strong electron-phonon coupling.~\cite{TT,Tsiovkin07} Thus, theoretical explanation 
of the high values of resistivity of transuranium metals will require significant efforts, 
and hence {\it ab-initio} calculations are of high interest 
for the understanding of ground state properties in these materials. 

Recent experiments and electronic structure calculations revealed numerous effects 
due to strong correlations of the 5$f$ electrons in Np, Pu, Am, and Cm.~\cite{Moore09} 
A number of band methods and approximations have been applied to describe magnetic 
and spectral properties of transuranium metals.~\cite{Moore09} 
Nonmagnetic ground state of pure plutonium metal observed experimentally\cite{Lashley05} 
was reproduced in the electronic structure calculations by the LDA+$U$+SO method.~\cite{Shorikov05} 
There the exchange interaction was found to be the reason of artificial antiferromagnetic 
ordering in previous LSDA+$U$ investigations. Also non-magnetic ground state was obtained 
in around-mean-field version of the LDA+$U$ method,~\cite{Shick05} then in LDA + Hubbard I 
approximation,~\cite{Shick09,Havela09} and hybrid density functionals with 
a dominant contribution of HF functional.~\cite{Atta09} Recently, the reliability 
of these results was proved by the more detailed analysis of exchange 
interaction,~\cite{Cricchio08,Bultmark09} and also by the LDA+DMFT method,~\cite{Kotliar06} 
combining the LDA approximation with the Dynamical Mean-Field Theory (DMFT) 
in various modifications.~\cite{Savrasov01,Pourovskii06,Pourovskii07,Shim07,Shim09,Zhu07,Marianetti08}
However, consistent interpretation of spectroscopic data by these calculations 
is not found yet.~\cite{Tobin08}

In this paper we report the results of electronic structure and resistivity calculations 
for transuranium metals at normal conditions and under pressure. In terms of CPA 
the many-band conductivity model was derived and applied for fcc-Pu, Am, and Cm 
and bcc-Np numerical investigations of electrical resistivity using DOSes of metals 
calculated within the LDA+$U$+SO method as a starting point. The derived model allows one 
to account for initial DOSes modifications in a direct way as a result of temperature erosion 
and its renormalization due to interband $s \rightarrow d$, $s \rightarrow f$, 
$d \rightarrow f$, and $f \rightarrow d$ electron transitions. The DOSes evolution 
of the metals with temperature at normal conditions and under pressure were presented 
and discussed. Results of the ER calculations are found in good agreement 
with available experimental data. In terms of the proposed conductivity model, 
using {\it ab initio} obtained parameters, the nature of high-resistivity 
values in actinides is also discussed.

\section{Electronic Structure Calculations}
\label{Method} 

We investigated the electronic structure of transuranium metals within 
the LDA+$U$+SO method described in detail in Ref.~\onlinecite{Shorikov05}. 
In this method the exchange interaction (spin polarization) term 
in the Hamiltonian is implemented in the general nondiagonal matrix form 
regarding the spin variables. This form is necessary for correct description 
of 5$f$ electrons for the case of $jj$ and intermediate couplings.~\cite{Shorikov05} 
The LDA+$U$+SO method is introduced by the effective single-particle Hamiltonian 
which supplements initial local density approximation (LDA) functional by 
a term with orbital dependent potential $V_{mm^{\prime }}^{ss^{\prime }}$, which accounts 
for Coulomb interactions,~\cite{LDA+U} and spin-orbit coupling term:~\cite{Shorikov05}
\begin{widetext}
\begin{equation}
\widehat{H}_{LDA+U+SO}=\widehat{H}_{LDA}+\sum_{ms,m^{\prime
}s^{\prime }}\mid inlms \rangle V_{mm^{\prime }}^{ss^{\prime
}}\langle inlm^{\prime }s^{\prime }|
+\lambda \cdot  \widehat{\mathbf L} \cdot \widehat{\mathbf S},
\label{Hamilt}
\end{equation}
\begin{equation}
\begin{array}{c}
V_{mm^{\prime }}^{ss^{\prime }}=\delta _{ss^{\prime
}}\sum_{m^{\prime \prime},m^{\prime \prime \prime}}\{\langle m,m^{\prime \prime }\mid V_{ee}\mid m^{\prime
},m^{\prime \prime \prime }\rangle n_{m^{\prime \prime }m^{\prime \prime
\prime }}^{-s-s } 
+(\langle m,m^{\prime \prime }\mid V_{ee}\mid m^{\prime },m^{\prime \prime \prime }\rangle
-\langle m,m^{\prime \prime }\mid V_{ee}\mid m^{\prime \prime \prime },
m^{\prime }\rangle )n_{m^{\prime \prime }m^{\prime \prime \prime }}^{ss}\} \\
\\
-\left( 1-\delta _{ss^{\prime }}\right) \sum_{m^{\prime \prime},m^{\prime \prime \prime}}\langle
m,m^{\prime \prime }\mid V_{ee}\mid m^{\prime \prime \prime },m^{\prime
}\rangle n_{m^{\prime \prime }m^{\prime \prime \prime }}^{s^{\prime
}s} -U(N-\frac{1}{2})+\frac{1}{2}J_H(N-1).
\end{array}
\label{Pot}
\end{equation}
\end{widetext} 

Here $U$ and $J_H$ are screened Coulomb and Hund exchange parameters 
which are determined in the constrain LDA calculations.~\cite{Gunnarsson89,Anisimov91a} 
The screened Coulomb interaction matrix elements 
$\langle m,m^{\prime \prime }\mid V_{ee}\mid m^{\prime },m^{\prime \prime \prime }\rangle$ 
could be expressed via these parameters.~\cite{LDA+U} In the constrain LDA procedure 
a {\it screened} Coulomb interaction of 5$f$ electrons is evaluated 
that requires the choice of screening channels taken into account in the constrain 
LDA calculations. Taking $s$, $p$, and $d$ channels, for Np, Pu, Am, and Cm 
we obtained the Coulomb parameter value $U \approx $ 4~eV in good agreement 
with the previously used value.~\cite{Savrasov01} 

In contrast to the direct Coulomb parameter $U$, the exchange parameter $J_H$ 
is evaluated as a {\it{difference}} of interaction energy for the electrons 
pairs with the opposite and the same spin directions. 
Here the screening process is defined by the charge but not spin state of the ion,
then the screening contribution is canceled for exchange Coulomb interaction 
and parameter $J_H$ does not depend on the choice of screening channels. 
For neptunium and plutonium, the value of Hund exchange parameter $J_H$ 
was reported to be $J_H$~=~0.48~eV,~\cite{Shorikov05} for americium -- 0.49~eV, 
and for curium -- 0.52 eV.~\cite{Tsiovkin07,LMTO} 

In Eq. (\ref{Pot}) the off-diagonal exchange interaction terms with 
$n_{m^{\prime \prime }m^{\prime \prime \prime }}^{s^{\prime }s}$ 
are significant for actinide elements. It was demonstrated 
in Ref.~\onlinecite{Shorikov05}, that omission of these terms can result in 
incorrect antiferromagnetic ground state for fcc-plutonium metal. 
This fact is due to the intermediate or $jj$ coupling scheme taking place 
in most actinide metals, including Pu, Np, and Cm. But in pure Am metal
$jj$ coupling is present. It means that total moment $\mathbf J$ 
is well defined, but not spin $\mathbf S$ and orbital $\mathbf L$ moments, 
as in usual $LS$ coupling scheme. In this case, the basis of eigenfunctions 
of total moment operator $\{jm_j\}$ is the best choice. The matrix 
of spin-orbit coupling operator is diagonal in this basis but not 
the exchange interaction (spin-polarization) term in the Hamiltonian.

Whereas in the $LS$ coupling scheme $\mathbf S$ and $\mathbf L$ operators 
are well defined. Then the basis of $LS$ orbitals, which are eigenfunctions 
of both spin $\mathbf S$ and orbital moment $\mathbf L$ operators, is a good choice. 
In this case it is possible to define quantization axis in the direction of spin
moment vector so that occupation and potential matrices will be
diagonal in spin variables. 

When the intermediate coupling is realized, neither $jj$ basis, nor $LS$ 
is valid, and occupation matrix is nondiagonal in both orbital bases 
and both terms in the Hamiltonian: spin-orbit coupling and exchange 
interaction, must be taken in a general nondiagonal matrix form, 
and some finer treatment is necessary. 
Intermediate basis is still cumbersome and is used for model atomic calculations.

For the actinide metals under consideration the LDA+$U$+SO densities of states (DOS) 
were taken as a starting point for further CPA simulations of ER temperature 
dependencies within the model described below.

\section{Coherent Potential Approximation for Many Band Conductivity Model}
\label{CPA} 

In the general case, the {\it many-band} conductivity model 
is suitable for explanation of kinetic properties in actinides 
due to equal probabilities of transitions of $(s+p)$ electrons into almost 
empty $d$ and $f$ bands, since the values of DOSes at the Fermi level 
in $d$ and $f$ bands are close to each other. Thus, ratios 
between partial $d$ and $f$ DOSes at the Fermi level lead to opening 
of additional channel of direct $d \rightarrow f$ and back $f \rightarrow d$ 
electrons transition. Moreover, the evolution of the $d$ and $f$ DOSes 
with temperature and direct and back transitions depends on the evolution 
of DOSes of the other shell. Then it is reasonable to obtain DOSes 
at finite temperatures with regular self-consistent procedure and 
derive corresponding CPA set of equations without model simplification, 
in the same way as it was made previously in Ref.~\onlinecite{TT}, when only direct 
$s \rightarrow d$ and $s \rightarrow f$ transitions were accounted for.

We start considering the $s$($p$), $d$, and $f$ electrons performing intra- 
and inter-band transitions as a result of their scattering at long wave phonons. 
We assume also that the accepting $d$ or $f$ bands are partially filled. 
Then the Hamiltonian of electron subsystem $\hat{H} = \hat{H}_0+\hat{H}_{int}$ 
can be written down in the following form: 
\begin{equation}
\hat{H} = \sum_{l} E_{l}\hat{a}^{+}_{l} \hat{a}_{l}
+\frac{1}{N}\sum_{n,l,l'}e^{-i(\vec{k}-\vec{k'}, \vec{R}_n)}
\hat{V}_{ll'}(n)\hat{a}^{+}_{l} \hat{a}_{l'},
 \end{equation}
where $E_{l}$ is a periodical part of electrons energy. Combined index
$l$ includes the band index $j$ $(j=s,d,f)$ and wave vector $\vec{k}$;
$\vec{R_{n}}$ is a radius-vector of the $n$-th site of a crystal
lattice.  Operator $\hat
{V}_{l,l'}(n)$ describes the intensity of
electron-phonon interaction. If thermal displacements of
ions are small, the operator $\hat {V}_{l,l'}(n)$ can be written as
 \begin{equation}
\hat{V}_{\alpha,l,l'}(n)=Z_{ll'} \frac{-i}{\sqrt{N}}
\sum_{\vec{q}} \sqrt{ \frac{q_0}{q}} \left[
e^{i(\vec{q}\vec{R_n})} \hat{ b}_{\vec{q}}
-e^{-i(\vec{q}\vec{R_n})} \hat{ b}^{+}_{\vec{q}}\right] , \label{F}
\end{equation} where
\begin{equation}\label{zz}
Z_{ll'}=\left( \frac{\hslash K_F}{2 M_{}S_{}}\right) ^{1/2} \cdot \left( \frac{2
K_F}{3q_0}\right)^{1/2} \cdot \Lambda_{l,l'}
\end{equation}
is the parameter of intensity of intra- and inter-band transitions due to electron-phonon scattering.
$ M$ and $ S$ are  the mass and the sound velocity in metal, respectively;
$q_0$ is the maximum value of $q$ and $K_F$ is the Fermi wave number of electron; $\Lambda_{l,l'}$
are the (Bloch) parameters of electron-phonon coupling.

Using {\it ab-initio} obtained DOSes of metal as a starting  point  of numerical 
calculations  we assume that  effects  of $s$, $p$, $d$, $f$ hybridization 
are accounted for in the electron ground state. For simplicity we keep 
the same band notations after renormalization.

A set of CPA equations can easily be derived using Dyson equation 
and definition of T-matrix. Let us determine the total resolvent 
of the full energy operator $\hat{H}$:
\begin{eqnarray} \hat{R}=(z-\hat{H})^{-1}, \end{eqnarray}
and the strictly diagonal part of the total resolvent $\hat{R}$
in the $\hat{H}_{0}$ representation:
\begin{eqnarray}
\hat{G}=(z-\hat{H_{0}}-\hat{\Delta})^{-1},
\end{eqnarray}
where $\hat\Delta$ is a strictly diagonal shift operator in the $\hat{H}_{0}$
representation.
Broadening of single electron levels is described as: 
\begin{eqnarray}
\hat{\Delta}\label{del}
=\frac{1}{N}\sum_{n,l,l'}e^{-i(\vec{k}-\vec{k'},\vec{R}_{n})}\Delta_{j}\delta_{jj'}\hat{a}^{+}_{l}\hat{a}^{}_{l}.
\end{eqnarray}
The real part of the coherent potential $\Delta_{j}$
determines the shift $\eta_{j}$, and its imaginary part
$\gamma ^{ }_{j}$ determines the broadening of
single-electron levels.

To derive equations within many-band CPA, the Dyson identity was used:
\begin{eqnarray}
\hat{R}=\hat{G}+\hat{G}(\hat{V}-\hat{\Delta})\hat{R}.\label{D}
\end{eqnarray}
Scattering operator $\hat{T}$ can be determined in a convenient form as:
\begin{eqnarray}
\hat{R}-\hat{G}=\hat{G}\hat{T}\hat{G}.\label{T}
\end{eqnarray}
Multiplying both parts of identity (\ref{T}) by $\hat{G}^{-1}$, one obtains 
the following expression for the scattering operator:
\begin{eqnarray}
\hat{T}=\hat{G}^{-1}(\hat{R}-\hat{G})\hat{G}^{-1}.\label{Dif}
\end{eqnarray}
Using the Dyson identity (\ref{D}) and Eq. (\ref{Dif}) for the
shift operator,  the following operator series can be written:
\begin{eqnarray}
\hat{\Delta}=[(\hat{V}-\hat{\Delta})\hat{G}(\hat{V}-\hat{\Delta})+
\nonumber \\
(\hat{V}-\hat{\Delta}) \hat{G} (\hat{V}-\hat{\Delta})
\hat{G}(\hat{V}-\hat{\Delta})+\ldots ]_{diag}.\label{Row}
\end{eqnarray}
Here the \emph{diag} index means that the diagonal part in the $\hat{H}_{0}$
representation of the sum of operator products in brackets should be taken.
The series (\ref{Row}) contains mutually compensated block terms, in which the shift
operator is included in indirect form.~\cite{VV} Excluding the compensated
block terms from Eq.~(\ref{Row}), and averaging over ion displacements, 
one finally obtains the shift operator as
\begin{eqnarray} \left\langle \hat{\Delta}\right\rangle =\left\langle \left[\hat{V}\hat{G}\hat{V}+
\hat{V}\hat{G}\hat{V}\hat{G}\hat{V}+...\right]_{D}\right\rangle .\label{Del}
\end{eqnarray}
Here only strictly diagonal terms in the $\hat{H}_{0}$ representation ($[...]_{D}$) 
are accounted for, and items containing the blocks are omitted. 
The brackets $<...>$ mean averaging over phonons. Let us assume also, 
that at high temperature the operator $\hat{V}_{ll'}(u)$ can be replaced 
with the one averaged over wave vectors $\vec{k}$ and $\vec{k'}$ functions 
of fluctuating variable $u$ -- $V_{n,jj'}(u)$.~\cite{chen} Then the series (\ref{Del}) 
can be summed up accurately (in the convergence range \mbox{$|V_j(u) F_j |<1$)} 
as it is shown in the Appendix for single-electron and single-site approaches. 
Using for simplification also \mbox{$|V_{n,sj}(u) F_{s}|  \ll 1 $}, 
\mbox{$\left|V_{n,s}F_{s}\right| /\left|V_{n,(d)f}F_{(d)f}\right| \ll 1$},
\mbox{$\left|V_{sj}F_{s}\right| /\left|V_{jj'}F_{j}\right| \ll 1$} 
and  notation $ V_{n,jj}(u)= V_{n,j}(u)$, for the averaging over ion thermal 
displacements $s$- and $d$-band coherent potentials, one obtains:
\begin{widetext}
\begin{eqnarray} \label{Eq1}\Delta^{}_{s}=\sum_{j\neq j'}   
\int_{-\infty}^{\infty}d u P(u)\frac{V^{2}_{j}(u)F^{}_{j}
[1-V_{j'}(u)F^{}_{j'}]+V_{sd}(u) V_{sf}(u)V_{df}(u)F^{}_{j}F^{}_{j'}}
 {[1-V_{d}(u)F^{}_{d}][1-V_{f}(u)F^{}_{f}]
-V_{df}^2(u)F^{}_{d}F^{}_{f}},
\end{eqnarray}
\begin{eqnarray}\label{EqCPA} \Delta^{}_{d}=   
\int_{-\infty}^{\infty}du P(u)\frac{V^{2}_{d}(u)F^{}_{d}
[1-V_{f}(u)F^{}_{f}]+V^2_{df}(u) F^{}_{f} [1+V_{d}(u)F^{}_{d}]}
 {[1
-V_{j'}(u)F^{}_{d}][1-V_{f}(u)F^{}_{f}]
-V_{df}^2(u)F^{}_{d}F^{}_{f}}.\end{eqnarray}
\end{widetext}
Where 
\begin{equation}
P(u)= \frac{1}{\sqrt{2 \pi \beta}}
e^{-u^2/2\beta},~~~~   \beta= Z_{\alpha,jj'} 2T/\theta_D
\end{equation} 
is the Gauss distribution function\cite{chen,Tsiovkin07} 
and $\theta_D$ is the Debye temperature. Equation for coherent potential 
of the $f$ band electrons is the same as Eq.~(\ref{EqCPA}) 
but a replacement of band indices $s\rightleftarrows f$ is necessary.

Note also, that an assumption $|V_{n,sj}(u) F_{s}|  \ll 1 $ 
in Eqs.~(\ref{Eq1}) and (\ref{EqCPA}) means physically that 
the partial filling of $d(f)$-bands sets the traps for mobile 
$s$-electrons and corresponds to the Mott idea, successfully 
applied previously for two band conductivity model. 

On the other hand, Eq.~(\ref{EqCPA}) describes direct and back 
transitions of $d$($f$)-electrons and provides a possibility 
of accounting for $d$($f$)-bands modifications with temperature. 
These equations are solved within self-consistent loop. 
A solution of these equations allows one to calculate coherent 
potential for $s$-conductivity electrons.

\begin{figure*}[!t]
 \begin{center}
  \epsfxsize=14cm
 \epsfbox{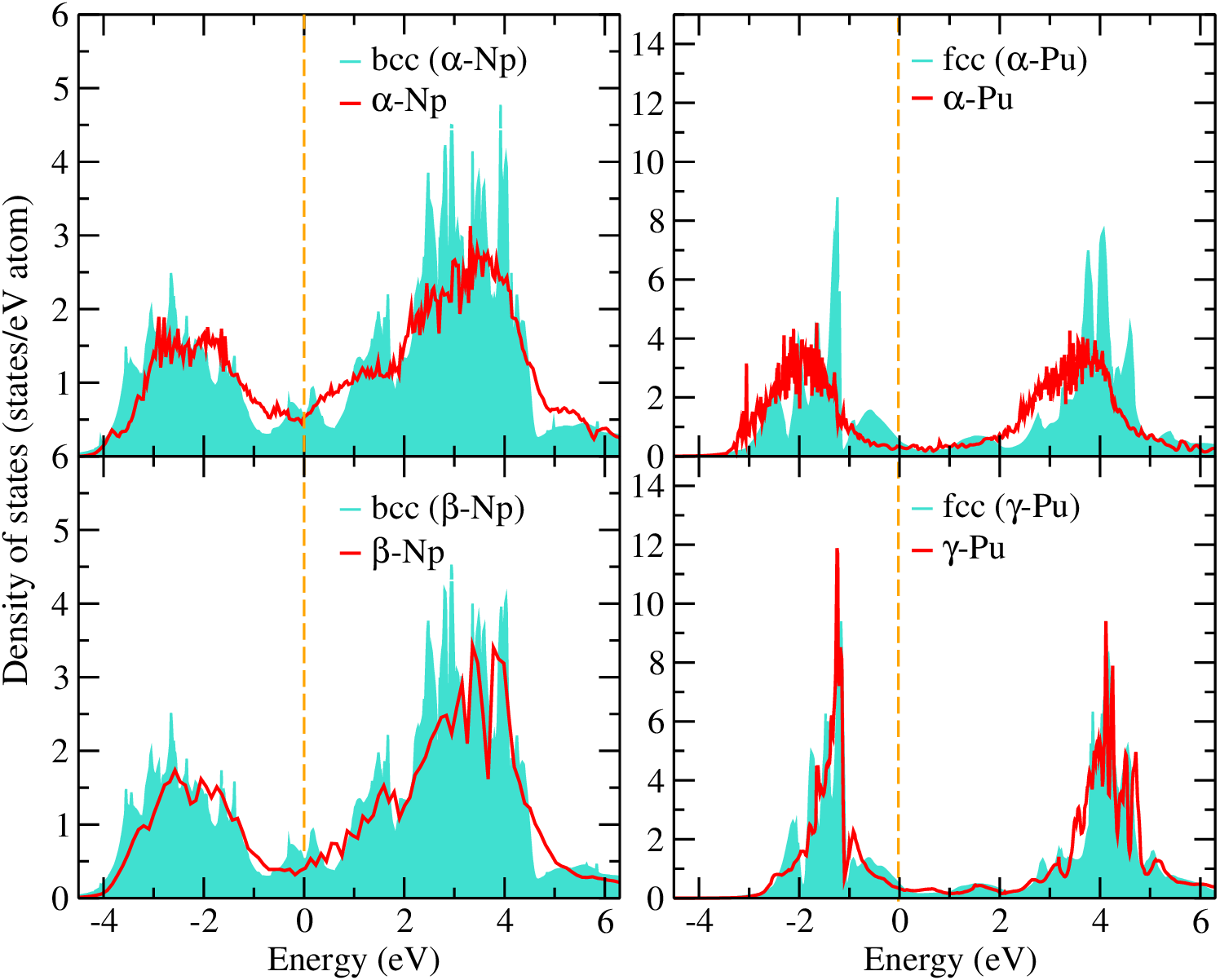}
 \end{center}
 \caption {(Color online) Partial 5$f$ densities of states for $\alpha$- and $\beta$-Np (left panels) 
 and $\alpha$- and $\gamma$-Pu (right panels) in real crystal structures 
 are compared with the bcc-(Np) and fcc-(Pu) DOSes for the corresponding cell volumes per actinide atom (see 
 Ref.~\onlinecite{Lukoyanov10}). The Fermi energy corresponds to zero.} 
 \label{fig:1}
\end{figure*}

\begin{table*}[!t]
\caption {Symmetry groups of Np, Pu, Am, and Cm and results of the electronic structure calculations 
for the real crystal structures within the LDA+$U$+SO method. The largest values of occupation matrices 
off-diagonal elements (OD) in $\{LS\}$ and $\{jm_j\}$ basis sets are given 
(see details in Ref.~\onlinecite{Shorikov05}). Then the calculated values for spin ($S$), 
orbital ($L$), total ($J$) moments, Land\`e factor, and effective magnetic moment 
($\mu^{calc}_{eff}$) in $\mu_{B}$ are presented.~\cite{noteM}}
\begin{center}
\begin{tabular}{ccccccccccc}
\hline \hline
Phase & Structure & OD$_{\{LS\}}$ & OD$_{\{jmj\}}$ & $n_{5/2}$ & $n_{7/2}$ & $S$ & $L$ & $J$ & $g^{calc}$ & $\mu^{calc}_{eff}$,~$\mu_{B}$ \\
\hline
$\alpha$-Np & Pnma\cite{Zah1}      & 0.29 & 0.27 & 3.04 & 1.11 & 1.28 & 4.25 & 2.97 & 0.68 & 2.33 \\
$\beta$-Np  & P42$_12$\cite{Zah2}  & 0.33 & 0.31 & 3.12 & 1.20 & 1.33 & 4.46 & 3.13 & 0.68 & 2.44 \\
$\gamma$-Np & Im3m\cite{Pearson67} & 0.36 & 0.33 & 3.13 & 1.14 & 1.40 & 4.68 & 3.28 & 0.67 & 2.53 \\
\hline
$\alpha$-Pu & P2$_1$/m\cite{Zachariasen55} & 0.41 & 0.25 & 4.41 & 1.68 & 0 & 0 & 0 & 0 & 0 \\
$\gamma$-Pu & C2/m\cite{Zachariasen55}    & 0.43 & 0.28 & 4.28 & 1.45 & 0 & 0 & 0 & 0 & 0 \\
$\delta$-Pu & Fm3m\cite{Zachariasen55}     & 0.45 & 0.01 & 5.57 & 0.24 & 0 & 0 & 0 & 0 & 0 \\
\hline
AmI         & P6$_3$/mmc\cite{Heathman01} & 0.47 & 0.30 & 4.65 & 1.62 & 0 & 0 & 0 & 0 & 0 \\
AmII        & Fm3m\cite{Heathman01}       & 0.47 & 0.02 & 5.89 & 0.52 & 0 & 0 & 0 & 0 & 0 \\
AmIII       & Fddd\cite{Heathman01}       & 0.45 & 0.24 & 3.91 & 2.43 & 0 & 0 & 0 & 0 & 0 \\
AmIV        & Pnma\cite{Heathman01}       & 0.44 & 0.27 & 4.55 & 1.71 & 0 & 0 & 0 & 0 & 0 \\
\hline
CmI         & P6$_3$/mmc\cite{Heathman05} & 0.33 & 0.51 & 4.71 & 2.88 & 2.85 & 0.91 & 3.76 & 1.76 & 7.44 \\
CmII        & Fm3m\cite{Heathman05}       & 0.31 & 0.45 & 4.72 & 2.78 & 2.77 & 0.75 & 3.52 & 1.79 & 7.13 \\
CmIII       & C2/c\cite{Heathman05}       & 0.31 & 0.49 & 4.82 & 2.82 & 2.77 & 0.75 & 3.52 & 1.79 & 7.13 \\
CmIV        & Fddd\cite{Heathman05}       & 0.36 & 0.41 & 4.77 & 2.54 & 2.39 & 0.83 & 3.22 & 1.74 & 6.42 \\
\hline \hline
\end{tabular}
\end{center}
\label{tab:cry}
\end{table*} 

Simple assumption leads to the previously obtained results. 
Indeed, using for example $V_{jj'}=0$, i.e., neglecting 
the inter-band transitions, three independent equations corresponding 
to the single-band model of the CPA can be obtained 
from Eqs. (\ref{Eq1}) and (\ref{EqCPA}):~\cite{Tsiovkin07}
\begin{eqnarray}\left\langle \Delta_{j}\right\rangle 
 = \int\limits_{-\infty}^{+\infty}du~ 
P(u)\frac{ V_{j}^2(u)  F_j}
{1-F_f V_{j}(u) } .
\end{eqnarray}
Also, one can found the result for two band conductivity model 
and perturbation theory series for the coherent potential.

Accounting for this renormalization, and also for the values of DOSes of bands 
and their modification with the temperature, one can estimate the magnitude of ER 
in actinides. Note, that effects of irradiation will not be taken into account 
in this result.

\section{Results}\label{Results} 
\subsection{Ground state of metals at normal conditions and under pressure}

\begin{figure*}[!t]
 \begin{center}
  \epsfxsize=14cm
 \epsfbox{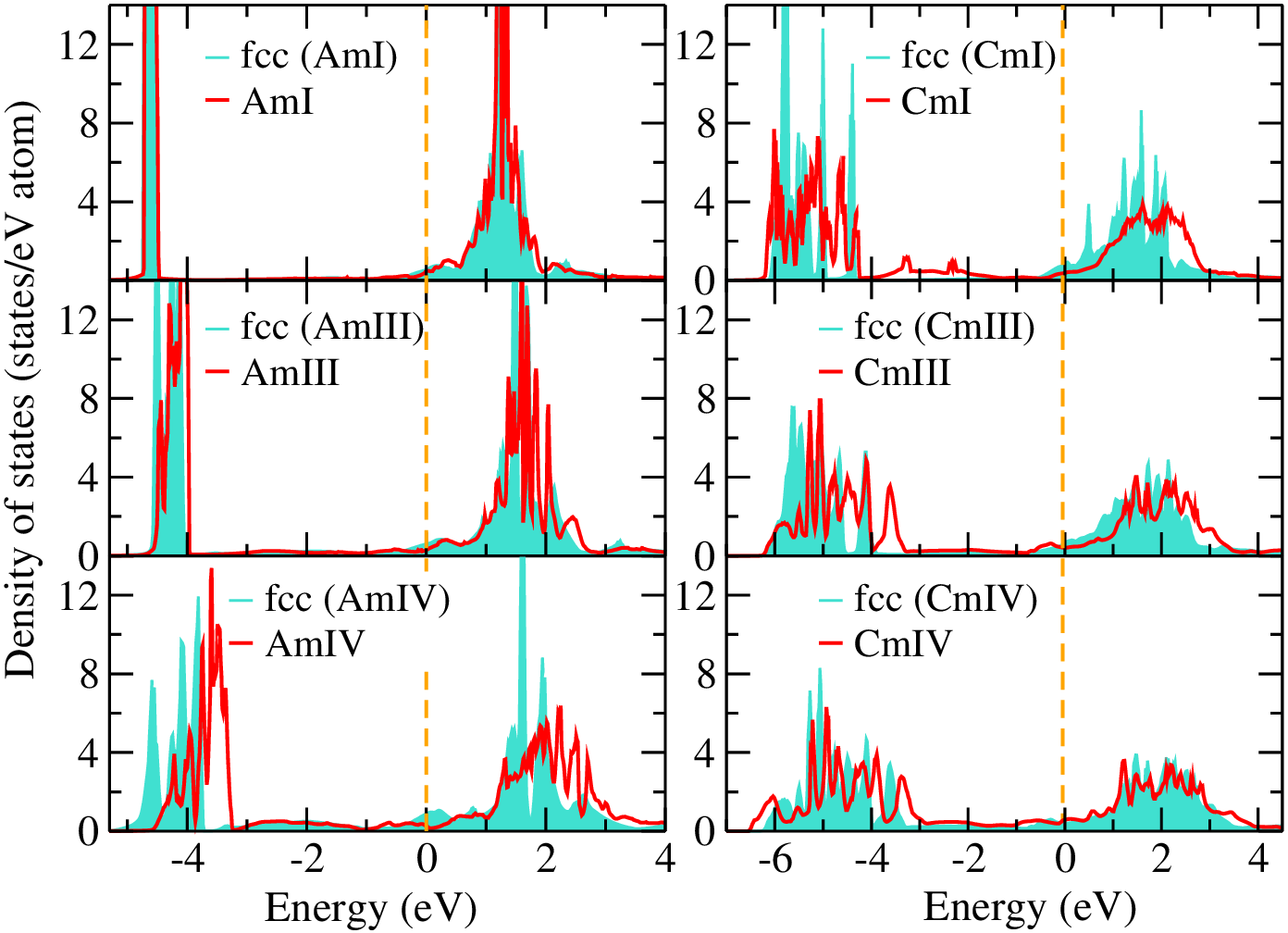}
 \end{center}
 \caption {(Color online) Partial 5$f$ densities of states for AmI, AmIII, and Am IV (left panels) 
 and CmI, CmIII, and Cm IV (right panels) in real crystal structures 
 are compared with the fcc-DOS for the corresponding cell volumes per actinide atom (see 
 Ref.~\onlinecite{Lukoyanov10}). The Fermi energy corresponds to zero.}
 \label{fig:2}
\end{figure*}

Electronic structure of Np, Pu, Am, and Cm in real crystal structures (Table 1) were calculated 
within the LDA+$U$+SO method. 
As one can see from the analysis of off-diagonal occupation matrix elements, in all 
metals (except $\delta$-Pu and AmII) in both LS and $jm_j$ bases off-diagonal elements 
are substantial and comparable. That means that intermediate coupling takes place in these metals. 
Simple cubic $\delta$-Pu and AmII are well described within the $jj$ coupling scheme that 
is confirmed by $f^6$ electronic configuration and smaller hybridization of $j$~=~5/2 subband
with $j$~=~7/2 subband and reflected in its smaller occupation. 

In all our calculations for Pu and Am only nonmagnetic solution was found, whereas in neptunium 
effective magnetic moment was obtained about 3 $\mu_B$ and depends on phase. In curium effective magnetic 
moment strongly differs in the phases, see Table 1. Experimental value of the moment in Cm 
was measured as 7.85 $\mu_B$.~\cite{Huray85} While model calculations predict magnetic moment 
from 7.94~$\mu_B$ in ionic picture in the assumption of pure $LS$ coupling to 7.6~$\mu_B$ 
in the assumption of intermediate coupling.~\cite{Huray85}  

For the many-band conductivity model presented in Section \ref{CPA} partial densities of states 
from $ab~initio$ calculations are used. In Figs.~\ref{fig:1} and \ref{fig:2} we present 5$f$ partial 
densities of states for the calculated crystal phases. Results for cubic ($\gamma$-Np, $\delta$-Pu, 
AmII, and CmII) structures under pressure were reported elsewhere.~\cite{Lukoyanov10}
In all DOSes one can distinguish two groups of peaks attributed to the subbands with 
the total moment value $j = 5/2$ at the lower energies and $j = 7/2$ at higher ones 
splitted by strong spin-orbit coupling. The Fermi level is shifted upward from the upper 
slope of $j = 5/2$ subband in Pu and crosses the $j = 7/2$ subband in Cm corresponding 
to the increasing number of $f$ electrons. A separation of the centers of gravity 
of these subbands for the value of Coulomb parameter $U$~=~4~eV results in 5 -- 5.5 eV. 

 \begin{figure}[!t]
  \begin{center}
   \epsfxsize=7cm
   \epsfbox{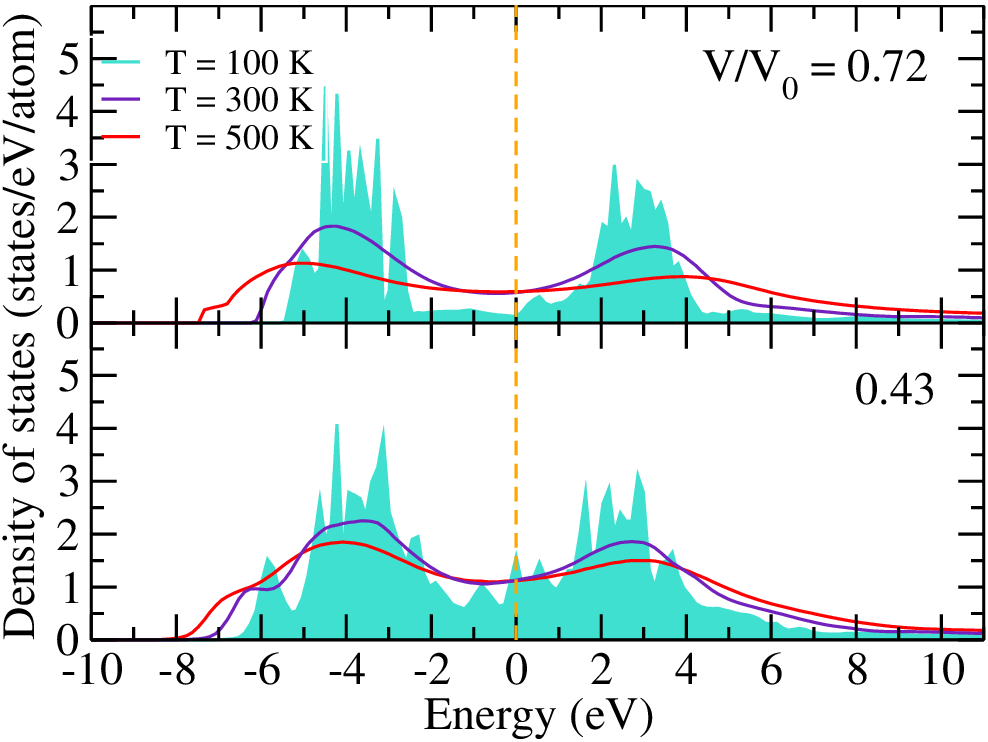}
  \end{center}
  \caption {(Color online) Partial 5$f$ densities of states of some fcc-curium volumes 
  under pressure the CPA broadened for the temperatures T~=~100, 300, and 500 K. 
  Here the volumes are related to V$_0$ -- the volume per ion in CmI phase at ambient pressure. 
  The Fermi energy corresponds to zero.}
  \label{fig:3}
 \end{figure}

For comparison we present DOSes of the cubic phase under pressure calculated within LDA+$U$+SO 
for the corresponding volumes per actinide atom. As one can see, cubic phases with the fitted volumes 
can be used as a good approximation for real phases 5$f$ DOSes, since the centers of gravity 
and bandwidth of $j$ = 5/2 and 7/2 subbands are found in good agreement with that in the real phases. 
Also the total density of states of real and corresponding cubic phases at the Fermi level 
are close. Having this close similarity in mind, below we report the resistivity model results 
for the DOSes of cubic phases with different volumes taken as a starting point. Since the model allows 
to estimate temperature dependence of ER, the starting DOSes were temperature broadened. 
The result of such broadening for curium in the volumes per atom, corresponding to real 
volumes are shown in Fig.~\ref{fig:3}. All other DOSes before broadening are reported 
in Ref.~\onlinecite{Lukoyanov10} and look similar with the broadening.

\subsection{Electrical resistivity temperature dependencies of metals 
at normal conditions and under pressure}

Electrical resistivity in CPA is usually calculated within  Kubo formula 
for the diagonal part of conductivity tensor 
\begin{equation}\label{kubo}\sigma = (\rho)^{-1}=\sum_{j}\frac{4 e^2 \hbar n_j}{3\pi^2 m^* } \int dU (-\frac{df}{dU}) \times \Upsilon (U)
\end{equation}
where
\begin{equation}
\Upsilon (U)= \int dE g_{j}(E)E \left[ \frac{\gamma^{}_{j}(U)}{(U-E-\eta_{j}(U))^2+\gamma^{2}_{j}(U)}\right]^2 ,
\end{equation}
and $g_{j} $ is DOS of the $j$-th conductivity band. 
In this equation, the approximate expression $v^2=2E/m^*$ 
for the square of electron velocity was used. 
Note, that apparent limitations of Eq.~(\ref{kubo}) arise 
from neglecting of back $f (d)\rightarrow s$ transitions 
of conductivity electrons.~\cite{Tsiovkin07} 

Numerical solution of Eqs. (\ref{Eq1}) and (\ref{EqCPA}) 
with $s,d$, and $f$ DOSes of metal provides a usual way to estimate 
effective mass of conductivity electrons ($m^*_j$) from  $ab~initio$ results. 
Accounting for effective mass renormalization and ``accepting''  DOSes modification 
with  temperature, one can try to estimate the absolute values of ER of actinides. 
However, this result will comprise no correction on effects of irradiation.

Full solution of the CPA Eq.~(\ref{EqCPA}) for pure fcc-Pu, Am, Cm and bcc-Np 
metals at normal conditions and under pressure were obtained using corresponding 
$ab~initio$ DOSes\cite{Lukoyanov10} as the starting point of iteration procedure. 
Bloch constant value equal to 0.8 E$_F $\cite{LashlyPRL} and experimental data 
for the Debye temperature\cite{LashlyPRL,Lashley05} $\theta_D \sim 100K $, 
velocity of sound and structural data (see Table 1) were used for 
the parametrization of these equations. Note, that the Debye temperature 
estimations\cite{LashlyPRL,Lashley05,TDD} presented previously 
for $\delta$-Pu by other authors differ significantly from each other. 
However, simulations  within self-consistent general thermodynamic model, 
accounting for effect of anharmonicity of lattice\cite {TDeb} 
gives the same value for Debye temperature as found experimentally 
in Ref.~\onlinecite{LashlyPRL}. 

It is well known that the value of DOS and its behavior in the vicinity 
of the Fermi level affect significantly kinetic properties of metal. 
All DOSes modifications with temperature and pressure has attracted 
a great deal of attention. The present results are based on the common 
numerical solution of a set of CPA equations (\ref{EqCPA}), 
performed for different pressure and temperature and 
demonstrate general trend  of strong influence  of  electron
phonon interaction on the initial  DOS.  
In Fig.~\ref{fig:3} the densities of states for a few volumes of curium 
are shown to illustrate the temperature broadening of DOSes. 
One can see from Fig.~\ref{fig:3}, 
that strong electron-phonon coupling leads to significant smoothing of all initial 
fine features of the DOS curves. At  the same time, applied pressure 
is slightly hindered the smoothing of the initial curve but does not lead 
to qualitatively different result.  In vicinity of the melting point, calculated 
DOSes of  all metals completely lose all their original features and  
are similar to each other. Calculated dynamics of DOSes evolution determines 
mainly the behavior of ER vs. temperature and applied pressure. 

Recently, ER calculations for pure bcc-Np, fcc-Pu, Am, and Cm metals within CPA 
for two band conductivity model at high temperatures and normal conditions 
proposed typical metallic behavior of ER over the whole temperature region without 
any specific peculiarities.~\cite{Tsiovkin07} Similar behavior of ER  vs. temperature 
without any anomalies and singularities were calculated in this work for actinides 
within the proposed conductivity model both at normal conditions and under pressure. 
For bcc-Np and fcc-Cm metal only the phonon part of ER was calculated. 
For pure fcc-Pu and Am at high temperature ER is defined by electron-phonon 
scattering mainly. 

One can see, that all theoretical curves of ER 
are similar to the common curves measured for a number of 4$d$-(5$d$-) 
transition  metals~\cite{AS} and agree well with the experimental data. 
Calculated dependencies of ER for all metals  show, that the TCR of the metals 
increases with the pressure and decrease with temperature for 100$\div $500~K. 
In the vicinity of melting point TCR for all metals was found to be a weakly 
increasing function of temperature. 

\begin{figure}[!t]
 \begin{center}
  \epsfxsize=7cm
  \epsfbox{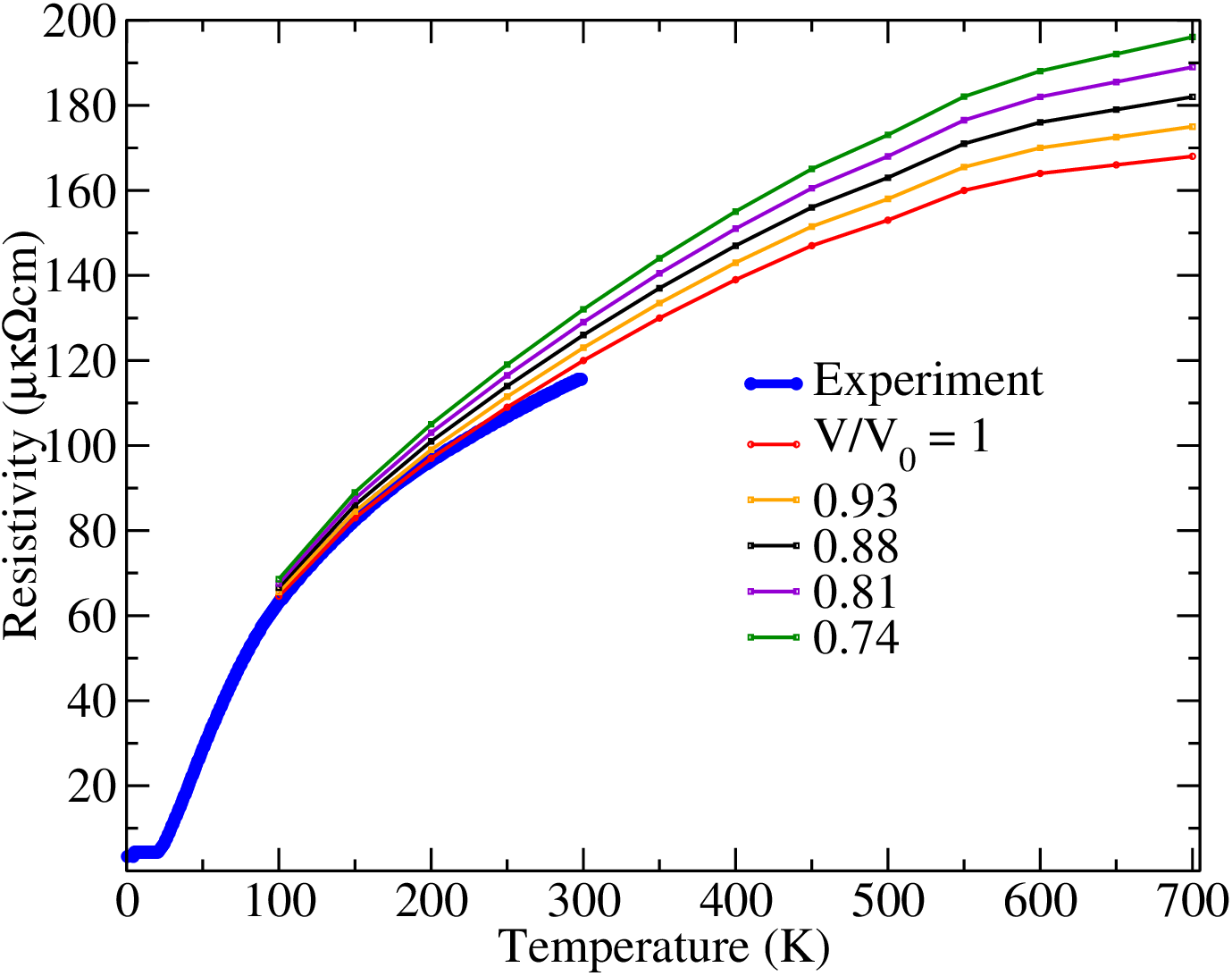}
 \end{center}
 \caption {(Color online) Electrical resistivity of Np metal under pressure. 
For each curve the corresponding volume is related to the volume per actinide ion 
(V$_0$) at ambient pressure -- V/V$_0$.}
 \label{fig:4}
\end{figure}

\textbf{Np metal.}~The first detailed  ER experimental data on temperature 
dependence of  pure orthorhombic $\alpha$-Np  phase with $R_{295}/R_{4.2}$=34.36 
was reported in 1963.~\cite{Np_res1} 
Subsequent experiments (see Ref.~\onlinecite{Oll}) were performed for polycrystalline 
samples of $\alpha$-Np in 4.2 -- 300~K temperature range. 
The residual resistivity of the Np metal was found to be equal 
to 12.2 $\mu\Omega cm$  and $R_{273}/R_{4.2}$ = 8.15,~\cite{Oll} that shows high 
concentration of different impurities in the samples. Authors 
of Ref.~\onlinecite{Np_res2} reported Np  $R_{273}/R_{4.2}$ value to be equal to 4.47 
and pointed out strong effect of self-damage on the observed value of ER. 

Note, that residual resistivity values about 10$\div $15~$\mu\Omega cm$ 
are typical in concentrated 3$d$- and 5$d$-transition metal alloys 
with strong electron impurity interaction. After subtraction of this 
background the temperature dependence of ER of Np was found to be ordinary 
for \textit{dilute alloys }with 
$\rho \sim T^{3} $ in low temperature region\cite{kom} and a weakly increasing  
non-linear function of temperature $T> \theta_D$. The calculated ER values 
agree well with experimental data in arbitrary units at normal conditions  
and predicts TCR increasing under pressure. Calculations of conductivity 
of $s$($p$)-, $d$-, and $f$-electrons show that total ER is determined 
by $s$($p$)-electrons mainly and partially by $d$-electrons. 
Assuming that $f$-band electrons are conducting ones, we obtained 
strongly overestimated ER values. Determining the temperature-dependent part of ER 
as $\rho(T)-\rho_0$ and using  calculated  values  of ER of $s$- and $d$-electrons, 
we found the resistivity values underestimated by 9--11~\%. 
That is why only electron-phonon interaction was taken into consideration 
and electron-electron coupling was neglected.

\begin{figure}[!t]
 \begin{center}
  \epsfxsize=7cm
  \epsfbox{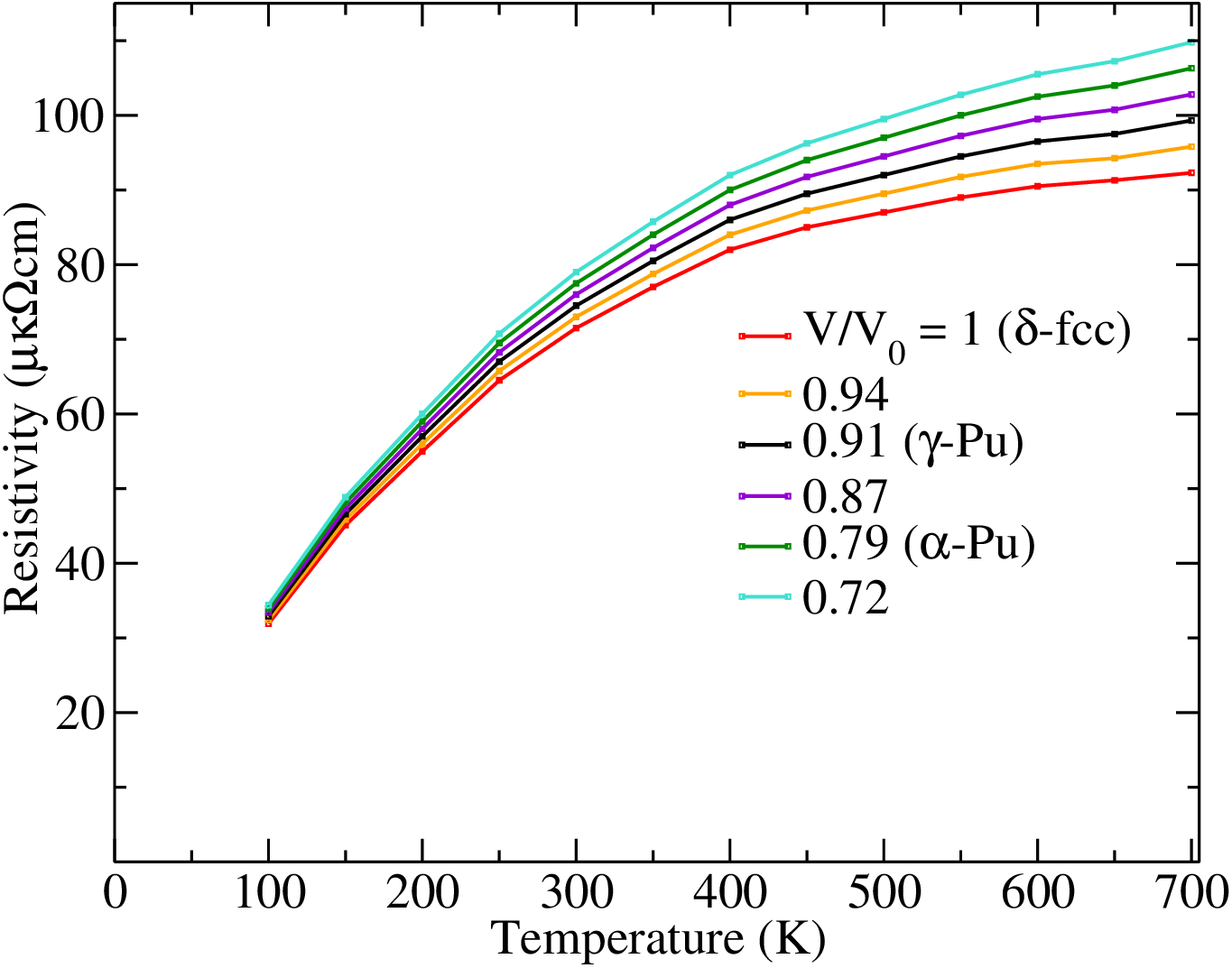}
 \end{center}
 \caption {(Color online) Electrical resistivity of Pu metal under pressure.}
 \label{fig:5}
\end{figure}

\textbf{Pu metal.} Negative TCR value in $\alpha$-Pu were measured about  
five decades ago.~\cite{Sm} This result is unique for pure metal and 
was confirmed by several groups.~\cite{Brod,Brodsky} Experimental results 
for ER motivated great discussion of possible mechanism of scattering 
which can provide such values of negative coefficient of resistivity 
and a model explaining it. 

However, note at first that $\alpha$-Pu 
as well as $\delta$-Pu have very high values of residual resistivity 
and high concentrations of impurities and defects. Strong change 
of ER values during the holding time also evidences for strong 
contributions of electron-impurity and electron-defect 
scattering parts to the total value of resistivity. 
Second, thus experimental data were obtained for dilute Pu-based 
alloys and quasi-polycrystalline samples. 

Using well-known experimental fact that ER 
of polycrystalline metals shows behavior analogous  
to their fcc- and bcc-crystal phases, the  negative  TCR  can easily be 
explained in terms  of electrons scattering on non-coherent phonons. 
Indeed,  electron scattering on lattice defects,  produced
by self-damage in  $\alpha$-Pu and ion mass modification as a result 
of irradiation leads to randomly distributed  ``impurity'' ions 
on the sites of crystal lattice. Hence, conductivity electrons 
scatter not only  on ``pure'' phonons  but also on randomly  distributed 
Coulomb fields of impurities ions. This model equivalence corresponds to interference model 
of scattering in fcc- dilute Pu-based  
alloys ($\delta$-Pu), previously proposed and discussed in Refs.~\onlinecite{Tsiovkin07} 
and~\onlinecite{TT}. This consideration can be extended, 
allowing a direct way to explain complex experimental data as well as non-magnetic 
nature of the observed ER  anomalies,  $\rho \sim T^2$  at low temperature, 
negative  TCR at high temperature range and exclude $\alpha$-Pu 
from list of anomalous metals, at least concerning its resistivity properties. 

Electrical resistivity calculations of fcc-Pu within proposed {\it many-band} conductivity model 
were performed in absolute units. Experimental ER data for pure fcc-Pu 
are unknown. Numerical simulation shows ordinary temperature dependence of ER at high temperature. 
An attempt to separate resistivity of different groups of electrons gives domination 
of $s$-band electrons with effective mass above 25$\div$30 at high pressure 
and $\sim$ 35$\div$40 at normal condition. The electron mass renormalization 
is the result of intensive $s \rightarrow d$ and $s \rightarrow f$ transitions 
and strong effect of direct $d \rightarrow f$ and back $f \rightarrow d$ 
transitions on the DOS values at the Fermi level. 
Calculated absolute values of ER underestimate experimental data 
for $\alpha$-Pu and $\delta$-Pu by a factor of 3 and predict 
metallic type of temperature dependencies of resistivity at normal 
conditions and under pressure. TCR of Pu was calculated as a weak 
increasing function of pressure and weak decreasing function of temperature.

\begin{figure}[!t]
 \begin{center}
  \epsfxsize=7cm
  \epsfbox{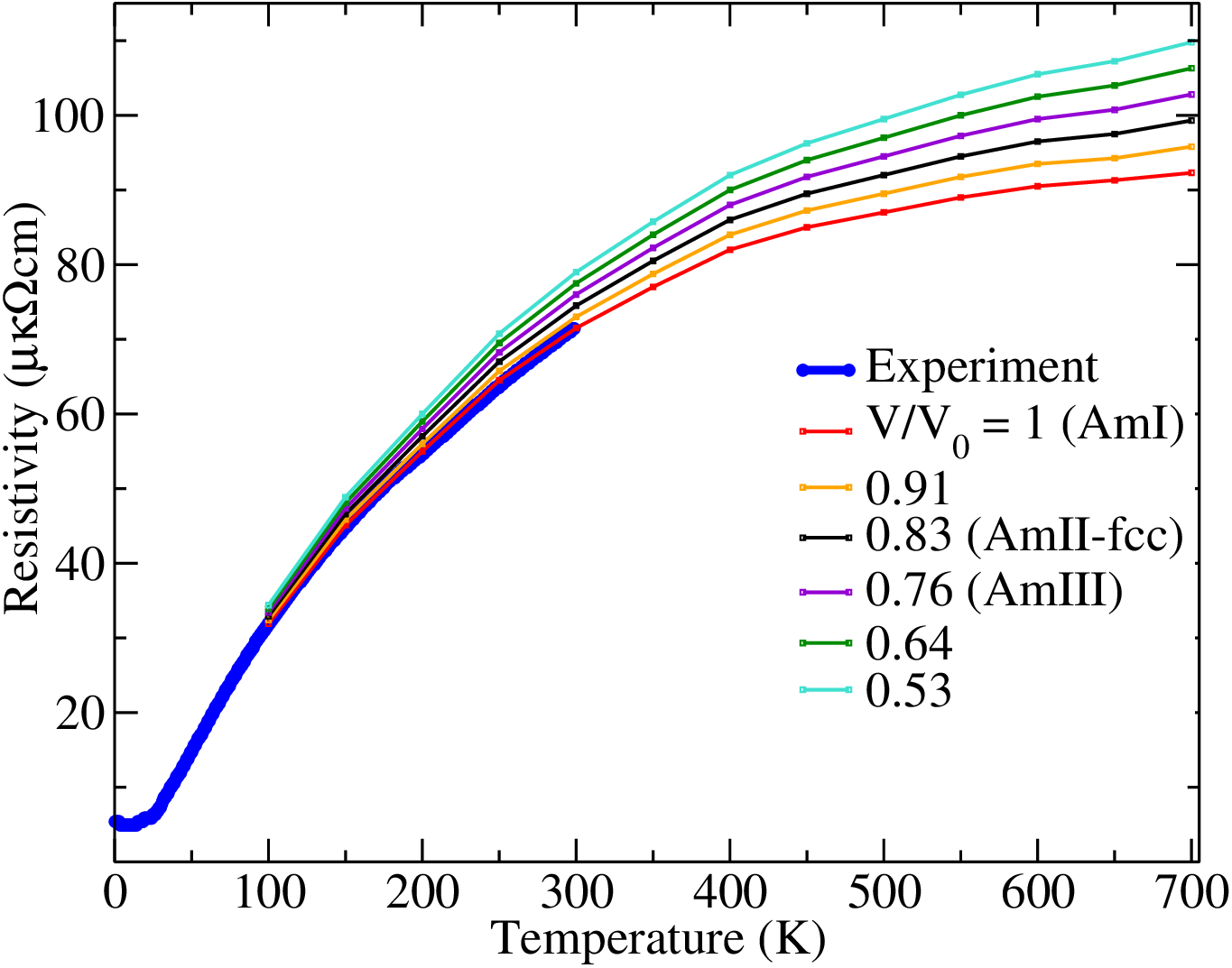}
 \end{center}
 \caption {(Color online) Electrical resistivity of Am metal under pressure.}
 \label{fig:6}
\end{figure}

\textbf{Am metal.}~During the last five years high interest in 
resistance properties of Am arose due to superconductivity found 
in pure metal. 
Reported values of $T_c$  of Am metal is above 0.5 K and found
to be dependent on applied pressure.~\cite{B} 

Electrical resistivity of pure Am  metal was reported for the first time in Ref.~\onlinecite{AmR}. 
The samples were relatively  pure  and low temperature behavior of 
Am metal  was obtained as $\rho(T)\sim T^{4\pm 0.5}$, that agrees well 
with the data for transition metals containing small number of impurities. 
At high temperature ER demonstrates also ordinary dependence 
with decreasing TCR for all investigated temperatures. 

Calculations performed within the proposed model at normal conditions and  pressure show
ordinary ER temperature dependencies  and agree well with experiments, see Fig.~\ref{fig:6}. 
Obtained TCR  is a decreasing function of temperature in  all regions but slightly 
increasing with applied pressure. Numerical values of ER  underestimate experimental 
data by 10 -- 15 \%. Nature of the obtained ER dependencies is analogous to calculated 
for Np and Pu ones and determined mainly by renormalization of DOSes with temperature.

At this time nothing can be say about giant values -- $\sim 400$~$\mu k \Omega cm$ of ER of Am-IV, obtained at room temperature and ambient pressure 25-30 GPa.~\cite{B} Moreover, these giant values of ER were measured in the system with typical metal type of conductivity without any sign of metal-semiconductor transition.  All possible model's  assumptions, accounting for different scattering mechanisms of conductivity of electrons and very optimistic expectations give ER values about 150-200 $ \mu k \Omega cm$ only.  The nature of high resistivity state can be explained assuming for example, that heavy electrons of $d$- and $f$-bands are the conductivity  electrons and $s$-type electrons are excluded from charge transport process. However, we have no corresponding experimental data to point out this opinion.

\begin{figure}[!t]
 \begin{center}
  \epsfxsize=7cm
  \epsfbox{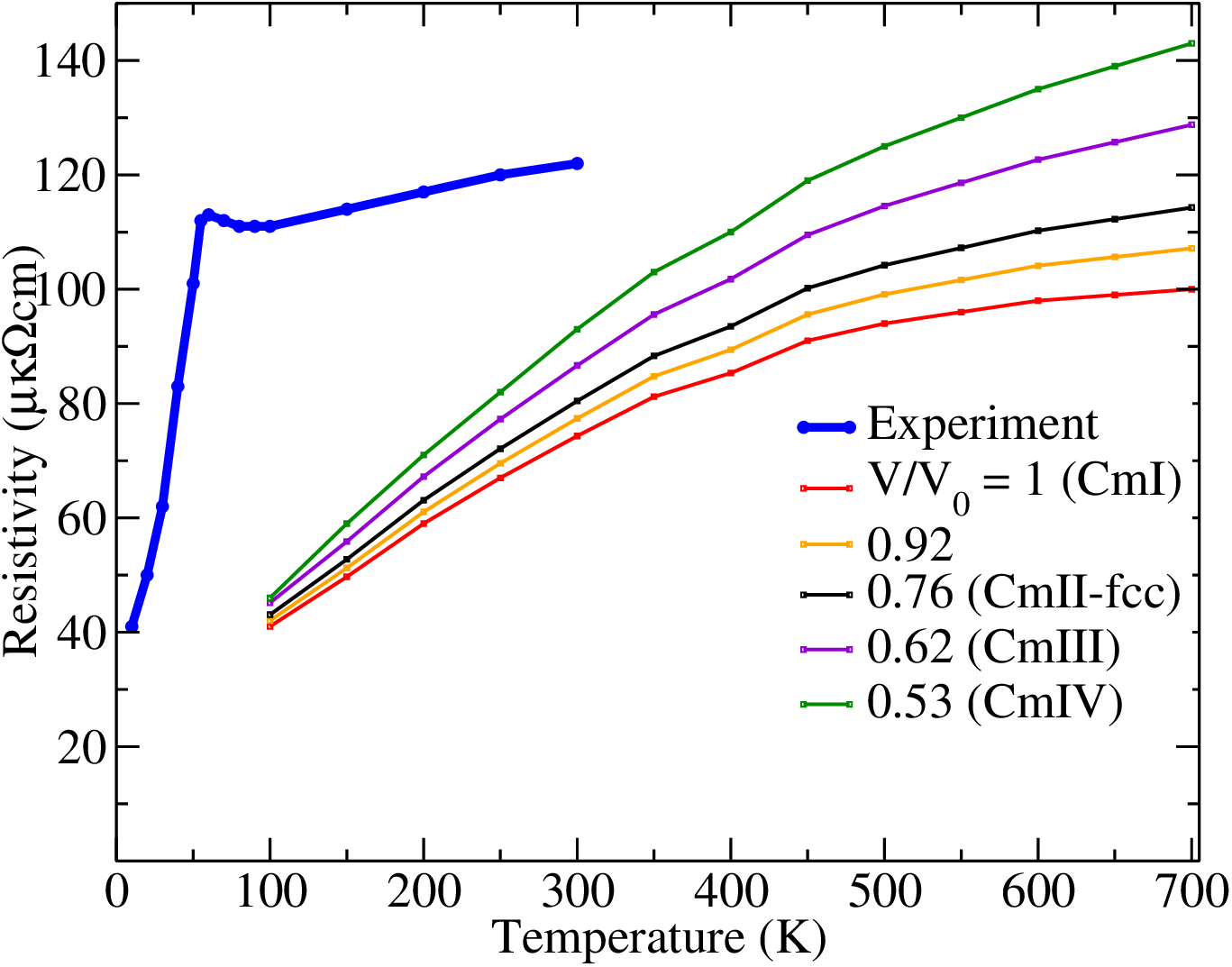}
 \end{center}
 \caption {(Color online) Electrical resistivity of Cm metal under pressure.}
 \label{fig:7}
\end{figure}

\textbf{Cm metal.}~Only recently experimental data on ER in $\alpha$-Cm (Cm-I) 
were reported. Total ER was found as a sum of residual resistivity 
(more than 40~$ \mu k \Omega \cdot cm$), magnetic part of resistivity, 
and the part of ER originated from electron-phonon coupling.~\cite{CmR} 
ER behavior of Cm is typical for antiferromagnetic metal and transparent anomalies 
in the vicinity of the Neel point. The magnetic part of ER  
was determined as a difference between the total resistivities 
of Cm and Am. Note first, that the large value of residual resistivity 
shows that the investigated samples of Cm metal contained a substantial number 
of impurities, including non-controlled ones. Second, nothing 
can be said about reliability of the suggestion, proposed in Ref.~\onlinecite{CmR},
that total ER is a sum of additive values, since strong
electron-impurity interaction  can significantly correct this result.  

In the present paper only the ``phonon'' part of the total ER in Cm 
was determined  and discussed unambiguously. The calculated ER 
is a usual linear function of temperature with positive TCR 
in the interval of 100~--~350~K and has similar behavior as in Pu 
and Am at the higher temperature. Temperature coefficient of resistivity 
of Cm demonstrates small decreasing under pressure, the similar one 
was calculated for the other metals under consideration. 
For this reason in Cm the DOS value at the Fermi level does not 
change drastically with temperature and mainly the $T/\Theta$ 
factor determines the temperature behavior of ER. Note, 
that from 500 to 700~K the ER of metals has weak temperature 
dependence and reaches its high temperature limit due to 
strong erosion of DOS at the Fermi level.

\section{Conclusion}
\label{Conclusion}

We report the results of model calculations of electrical resistivity 
for Np, Pu, Am, and Cm metals under pressure within CPA 
for many-bands conductivity model. 
We used DOSes of the metals under consideration 
as a starting point in our model calculations. For this purpose, 
Np, Pu, Am, and Cm pure metals were investigated in real crystal phases 
within the LDA+$U$ method with spin-orbit coupling (LDA+$U$+SO). 
For Am and Pu metals in all structures we found nonmagnetic ground state. 
We compared DOSes calculated for the real crystal structures of the transuranium 
metals with the DOSes for corresponding fcc volumes per actinide atom. 
It follows from this comparison that total density of states, as well as 
centers of gravity and bandwidth of $j$ = 5/2 and 7/2 subbands are in good agreement. 
It means that the cubic structures in different volumes 
provide good estimation for the DOSes of real crystal structures.

Results of our CPA calculations within derived many-band conductivity model 
for \textit{pure} fcc-Pu, Am, and Cm and bcc-Np 
using Bloch constant value $\sim 0.8 E_F$\cite{LashlyPRL} and  
{\it ab~initio} DOSes of metals show ordinary metal type 
of ER behavior vs. temperature  at normal conditions and under pressure. 
High values of ER in these metals are a consequence of $s \rightarrow d$ 
and $ s \rightarrow f$ transitions of scattering electrons and 
effective mass of conductivity electrons increasing as a result 
of $s$- and $d$-electron bands  hybridization. Weak non-linearity 
in the ER temperature behavior of metals above $\theta_D$ is caused 
by DOS erosion at the Fermi level due to electron-phonon interaction.

The obtained high-resistivity values in actinides result from $s \rightarrow d$ 
and $s \rightarrow f$ interband transitions and conductivity electron mass 
renormalization due to strong $s$-$d$ hybridization. Also strong influence on 
the ER  temperature dependencies of actinides is a result of DOS of accepting bands 
modifications as products of direct $s \rightarrow d$ and $d \rightarrow f$ 
and back $f \rightarrow d$ electrons transitions. Calculated phonon part 
of resistivity underestimates experimental data by a factor of 1.5 for Cm 
and by a factor of 3 for ideal Pu.

\section{Acknowledgments}
The authors are grateful to M.~Fluss and  L.~Timofeeva for stimulating conversations. 
This work was supported by the Russian Foundation for Basic Research, 
projects nos. 10-02-00046a, 09-02-00431a, and 10-02-00546a, 
Russian Federal Agency for Science and Innovations 
(Program ``Scientific and Scientific-Pedagogical Training of the Innovating 
Russia'' for 2009-2010 years), grant No. 02.740.11.0217,  
the Scientific program ``Development of scientific potential of universities'' 
no. 2.1.1/779. 

\section{Appendix}

Let us assume that the matrix elements of interaction $ V_ {ll'} $ are averaged over the angle 
between the wave vectors $ \vec{k} $ and $ \vec{k'} $ and thus depend only on 
the band index. Using this simplification and Eq. (\ref{Del}), one obtains:
\begin{widetext}\begin{eqnarray}
\hat{\Delta}=\frac{1}{N} \sum_{l n}
e^{i(\vec{k}-\vec{k'},\vec{R_n})} \left\{
\sum\limits_{j_1} [V_{jj_1}(u) ] F_{j_1} [V_{j_1j'}(u)]+
+ \sum\limits_{j_1 j_2} [V_{jj_1} (u)] F_{j_1} [V_{j_1j_2}(u) ] F_{j_2} [V_{j_2j'}(u)] + \cdots \right\} \delta_{ll'}a_l^{+} a_{l'},
\label{tjj}
\end{eqnarray}
where the $j$-th band electron Green function $F_{j}$ is defined as 
\begin{eqnarray}
F_{j}=\frac{1}{N}\sum_{{\vec{k}}}\frac{1}{(z-E_{\vec{k},j}-\Delta_{j})}.
\end{eqnarray}
Comparing Eqs. (\ref{del}) and (\ref{tjj}), for $\Delta_j$ one obtains
\begin{equation}
\Delta_{j'}= \sum\limits_{j_1} [V_{jj_1}(u) ] F_{j_1} [V_{j_1j'}(u) ] 
+ \sum\limits_{j_1 j_2} [V_{jj_1}(u) ] F_{j_1} [V_{j_1j_2}(u) ] F_{j_2} [V_{j_2j'}(u) ]  {+} \cdots .\label{row}
\end{equation}
Using the following matrix form 
\begin{eqnarray}
[F]=\left[\begin{array}{*{4}c} F_{s} \hfill & 0\hfill& 0\hfill\\
\hfill 0 & F_{d}\hfill & 0\hfill\\
\hfill 0 &  \hfill0& \hfill F_{f}\hfill\\
\end{array}\right],~
[\Delta]=\left[\begin{array}{*{4}c} \Delta_{s}& 0 \hfill& 0 \hfill\\
\hfill 0 & \Delta_{d} \hfill& 0 \hfill\\
\hfill 0 & 0 \hfill & \Delta_{f} \hfill\\
\end{array}\right],~~
[V]=\left[\begin{array}{*{4}c} V_{n,ss}(u) \hfill &V_{n,sd}(u) \hfill &V_{n,sf}(u) \hfill\\
 \hfill V_{n,ds}(u)\hfill & V_{n,dd}(u) \hfill & V_{n,df}(u)\\
 \hfill V_{n,fs}(u)\hfill & V_{n,fd}(u) \hfill & V_{n,ff}(u)\\  \end{array}
\right],
\end{eqnarray}
the series (\ref{Del}) is summed up accurately in the convergence range $|V_j(u) F_j |<1$.
Within single-electron and single-site approaches that gives 
\begin{equation}\label{ss}
\left\langle [\Delta]\right\rangle  = \left\langle [V][F][V]+[V][F][V][F][V]+\cdots\right\rangle 
=\left\langle \frac{[V][F][V]}{1-[F][V]}\right\rangle.
\end{equation}
Calculating matrix products and averaging over phonons both part of Eq.~(\ref{ss}), 
the set of CPA equations can be obtained.
\end{widetext}


\begin{thebibliography}{99}
\bibitem{remark} 
It is a well-known fact, that the order in $\rho \sim T^n$ dependencies  
increases up to n = 5 with purity of transition metals.
\bibitem{noteM}
For the phases with several inequivalent actinide ions we found close values 
of the off-diagonal elements and moments. For such phases the averaged values per ion 
and the largest OD elements are given.
\bibitem{kom} 
The dependence $T^{3.12}$ is more correct.~\cite{Np_res1}
\end{thebibliography}
\end {document}